# The Role of the Assumptions for the Existence of a General Equilibrium

Pablo Ahumada[*]

## Abstract

General Equilibrium Theory is the benchmark of economics, especially its results concerning the efficient allocation of resources, known as the First and Second Welfare Theorems. Yet, General Equilibrium Theory is beyond the scope of most economists. This paper is pitched as the first entry point into the theory. General Equilibrium Theory proves that at least one state of equilibrium always exists. In its most general approach, it uses fixed-point theorems to this end. This paper discusses the assumptions on individuals' behaviour and the structure of the system of exchange that guarantee that the conditions of the fixed-point theorems are satisfied. The purpose is to lay bare the role each plays in proving the existence of equilibrium and provide a clear picture of the relationship between the assumptions and the result. The discussion is presented in the simplest possible setting that captures the fundamental features of commodity exchange.

---

[*] Department of Economics, The University of Melbourne.

# 1. Introduction

General Equilibrium Theory and the concept of perfect competition underpinning this theory are the benchmark of economics, especially when it comes to optimality considerations. The First and Second Welfare Theorems of Economics, whereby every competitive equilibrium is optimal, and every optimal allocation could be achieved as a competitive equilibrium, are held in high regard to this day.[1] Hence the widespread belief that market competition is generally good and the setting to strive for, even if perfect competition is unattainable in practice. Despite the pervasiveness of this mindset in economics, General Equilibrium Theory, a highly theoretical field, is still beyond the scope of most economists, even in its most reader-friendly presentations.[2] The goal of this paper is to bridge this gap by offering a conceptual review of the problem of proving the existence of a general equilibrium. Specifically, this paper focuses on the most general assumptions that guarantee the existence of a general equilibrium and the role that each plays in ensuring this result.

Walras ([1874] 1954) pointed out that in a global system of interrelationships where everyone must support themselves, economic equilibrium is essentially a general equilibrium of exchange. However, the existence of equilibrium in such a setting remained only a conjecture until the 1930s. Proving the existence of equilibrium required a change of framework that entailed setting up the problem in terms of set theory and applying fixed-point theorems.[3] There are now four distinct approaches to proving the existence of a general equilibrium in a competitive framework.[4] The classical approach relies on the application of fixed-point theorems. This has remained the flagship theory and was spearheaded by Debreu and Arrow.[5] A second approach provides a constructive answer to the existence problem by developing algorithms to compute an approximate equilibrium.[6] A third approach relies on the theories of the degree and fixed-point index of a map.[7] Finally, the fourth approach relies on global analysis. It provides a constructive proof of existence through a differential process that converges to equilibrium.[8]

---

[1] For an exposition of both theorems, see Debreu (1959, Ch 6), Arrow and Hahn (1971, pp. 93-96) and Mas-Colell, Whinston, and Green (1995, Ch. 16).

[2] For excellent presentations of General Equilibrium Theory, see Debreu (1959, 1982), Arrow and Hahn (1971, Chs. 2-5) and Mas-Colell et al. (1995, Chs 15-17).

[3] See Arrow and Hahn (1971, ch. 1).

This paper focuses on the first approach. This approach is the most general as it can deal with correspondences to characterise both consumption and production.[9] Moreover, it only requires such correspondences to be upper hemicontinuous. The other strategies, in contrast, hinge on continuous functions that, for the most part, are differentiable.[10] Additionally, the paper is set in a pure exchange framework, which only includes exchange and consumption and leaves aside production, over a single period at a single location and with complete certainty. While production adds technical difficulties to proving the existence of an equilibrium, it adds nothing to the General Equilibrium framework from a conceptual viewpoint. The economic agents, both consumers and producers, are ultimately bounded by the total resources available and their initial allocation. These resources are in the hands of consumers, who own all firms, so all profits flow to them. The total resources are also all the potential wealth available in the economy, which appears in either its initial form or some processed form suitable for consumption.[11] As for the simplifications regarding the time horizon, location and uncertainty, Debreu (1959, Chs. 2 and 7) shows they can be introduced into the framework without formally changing the framework at all. Finally, the number of goods and agents is finite and all agents are price-takers and assumed to be rational.

---

[4] See Debreu (1982, pp. 697, 698).

[5] For examples of this approach, see Arrow and Debreu (1954), Debreu (1956, 1959, 1962, 1982), Arrow and Hahn (1971), McKenzie (1954, 1959), Gale (1955), Nikaidô (1956) and Uzawa (1962).

[6] For examples of the second approach, see Scarf (1982) and Mantel (1966).

[7] For examples of the third approach, see Balasko (2016) and Dierker (1982).

[8] For an example of the last approach, see Smale (1981).

[9] Correspondences are point-to-set functions, where each point of their domain returns a set of values rather than just one like true functions.

[10] See, for example, Debreu (1970), Smale (1981), Dierker (1982) and Balasko (2016).

The paper proceeds from the theorem used to prove the existence of equilibrium to the assumptions underlying the economic framework that guarantee the fulfilment of the theorem's assumptions, much in the style of Debreu (1982). However, the proof is not discussed as the paper focuses on the assumptions themselves. The paper does not aim for breadth but depth of understanding in the simplest possible setting that picks up all the essential features of commodity exchange. It comprises five sections. Section 2 presents the fundamental mathematical definitions and the theorem needed to prove the existence of equilibrium. Section 3 discusses the assumptions on the excess demand correspondence and the correspondence determining the relative prices of goods so that Kakutani's fixed-point theorem can be used to prove the existence of an equilibrium. Section 4 discusses the assumptions on individuals that underpin the assumptions discussed in Section 3. Section 5 concludes.

## 2. Mathematical Definitions and Main Result

This section presents the fundamental mathematical definitions and the main result needed to assess the assumptions required to prove the existence of a general equilibrium. It is based on Debreu (1959, ch. 1; 1982), Green and Heller (1981) and Fuente (1999) and is provided solely for ease of reference.

---

[11] For a detailed discussion of the production sector and how it is fitted into General Equilibrium Theory, see Debreu (1959, chs. 3 and 5) and Arrow and Hahn (1971, ch. 3).

Let $A$ and $B$ be two sets or collections of unique elements. We say that set $A$ is included in set $B$ or is a *subset* of $B$, denoted $A \subseteq B$, whenever every element of set $A$ is also an element of set $B$. That is, $A \subseteq B \Leftrightarrow (x \in A \Rightarrow x \in B)$. We say that set $A$ is a *strict subset* of $B$, denoted $A \subset B$, whenever $A \subseteq B$ but $B \nsubseteq A$. Put differently, there exists at least one element $x \in B$ such that $x \notin A$. Finally, sets $A$ and $B$ are *equal* if and only if set $A$ is a subset of $B$ and set $B$ is a subset of $A$. Formally, $A = B \Leftrightarrow (A \subseteq B \text{ and } B \subseteq A)$.

The *intersection* of sets $A$ and $B$, denoted by $A \cap B$, is the set of those elements that belong to both sets $A$ and $B$. Formally, $A \cap B = \{x | x \in A \text{ and } x \in B\}$. The *union* of sets $A$ and $B$, denoted by $A \cup B$, is the set of elements that belong to set $A$ or set $B$ or both. That is, $A \cup B = \{x | x \in A \text{ or } x \in B\}$. The *complement* of set $B$ in some set $A$ that contains set $B$, denoted by $B^C$, is the set of all points in set $A$ which do not belong to set $B$. Formally, $B^C = \{x \in A | x \notin B\}$. The *Cartesian product* of sets $A$ and $B$ is the set of all ordered pairs $(x, y)$ such that $x$ is in $A$ and $y$ is in $B$. That is, $A \times B = \{(x, y) | x \in A \text{ and } y \in B\}$.

Let $x$ denote a real number and $|x|$ its absolute value. A sequence of numbers $\{x_n\}$ such that for every positive integer $n \in N$, $x_n \in R$, tends to $x \in R$ if $x_n$ gets as close to $x$ as desired for $n$ large enough. Formally, for any number $\varepsilon > 0$ there exists a positive integer $n'$ such that for all $n > n'$, $|x_n - x| < \varepsilon$. Then, $x$ is the *limit of the sequence* $\{x_n\}$.

A *Euclidean space* is a finite-dimensional space over the real numbers. Let $l \in N$ be some finite natural number. A *limit point of a subset $B$ of the Euclidean space $R^l$*, $B \subseteq R^l$, is a point $x \in R^l$ such that there is a sequence of points $\{x_n\}$ in subset $B$ that tends to $x \in R^l$. For any number $\varepsilon > 0$ there exists a positive integer $n'$ such that for all $n > n'$, $|x_n^j - x^j| < \varepsilon$ for all coordinates $j = 1, \cdots, l$. Put differently, $x$ is a limit point of $B$ if there are points of $B$ arbitrarily close to $x$. The *closure* of set $B$ is the union of set $B$ and the set of all its limit points $\ell p(B)$. It is defined as $\bar{B} = B \cup \ell p(B)$. Set $B$ is *closed* if and only if it is equal to its closure, that is, if and only if $B = \bar{B}$. This means that the set has no holes.[12]

A point $x \in R^l$ can also be thought of as a *vector* made up of $l$ components. Consider $x_1, x_2 \in R^l$. We say $x_1 < x_2 \Leftrightarrow (x_1^j \leq x_2^j \ \forall j \ and \ \exists j \ such \ that \ x_1^j < x_2^j), j = 1, \cdots, l$. That is, one vector is *smaller* than another if and only if all its components are no larger than the other vector's, with at least one component being strictly smaller. We say one vector is *strictly smaller* than another if it is smaller in all its components, i.e. $x_1 \ll x_2 \Leftrightarrow (x_1^j < x_2^j \ \forall j)$. We say two vectors are *equal* if all their components are equal, i.e. $x_1 = x_2 \Leftrightarrow (x_1^j = x_2^j \ \forall j)$.

Let $S$ be some Euclidean space. A point $x \in S$ is a *boundary point* of set $B \subset S$ if and only if $x \in \bar{B} \cap \overline{B^C}$. Put differently, a boundary point of some set $B$ belongs to both its closure and that of its complement. The *boundary* of $B$, denoted by $\partial B$, is the set of all boundary points of $B$. A point $x \in S$ is *interior* to $B$ if and only if $x \notin \overline{B^C}$, that is, only if the point is not a limit point of the complement of set $B$ in the space $S$. The set of all points interior to $B$ is $\text{int} B = \left(\overline{B^C}\right)^C$. In a Euclidean space $S$, set $B$ is *bounded* if and only if it is strictly contained in some closed cube $K$, that is, if $B \subset K$. This means that the set has a boundary about it within the space in which it is contained.

Set $B \subset R^l$ is *compact* if and only if it is *closed* and *bounded*. This means it is a set with no holes and a boundary about it, which it contains, although it may be made up of separate pieces if it is not connected.

---

[12] Green and Heller (1981) define all concepts in this section for metric spaces in general. For the purpose of this paper, however, it is enough to define them for Euclidean spaces.

Set $B$ is *convex* if and only if it contains the segment joining any pair of points that belong to it. That is, set $B$ is convex if and only if $[x, y] \subseteq B$, for any pair of points $x, y \in B$. Every convex set is connected by definition. Therefore, a convex compact set is a bounded set with no holes comprised of a single piece. Moreover, every point in it is visible from every other.[13]

Consider two non-empty sets $A$ and $B$. A *function* $f$ from set $A$ into set $B$, denoted by $f : A \to B$, is a rule that maps each element $a \in A$ to one element $b \in B$. Set $A$ is called the *domain* of function $f$ and set $B$ is the *codomain* of $f$. Consider a subset $S \subseteq A$. $f(S) = \{f(s)|s \in S\}$ is called the *image* of set $S$ under the function $f$. It is the set of all points $\{b \in B\}$ to which function $f$ maps all points $\{s \in S\}$. $f(A)$ is the *range* of function $f$. Put differently, range($f$) is the image of its entire domain.

A *correspondence* is a point-to-set function. It associates each element of its domain with a unique non-empty set of elements. Let $A$ and $B$ be any two sets and $\varphi$ denote a correspondence from set $A$ into set $B$. Formally, $\varphi : A \to 2^B$, where $2^B$ denotes the set of all possible subsets of set $B$. Analogously to a function, a correspondence is written $Y = \varphi(x)$, where $Y$ indicates that the image of each point in the domain of $\varphi$ is a set of points. The *graph* of a correspondence $\varphi$ is the set of elements $(x, y)$ such that $y$ belongs to the image set of $x$ for each point $x$ in the domain of the correspondence. Formally, $G(\varphi) = \{(x, y) \in A \times B | y \in \varphi(x)\}$.

Let $A$ and $B$ be two subsets of two Euclidean spaces. A correspondence $\varphi$ from $A$ into $B$ is *upper hemicontinuous* at a point $x$ of $A$ if the following two conditions hold. There is a neighbourhood of $x$, $B_\varepsilon(x) \cap A$, in which $\varphi$ is bounded, where $B_\varepsilon(x) \cap A$ is the intersection between an open ball about point $x$ of radius $\varepsilon > 0$ and the domain of the correspondence $\varphi$. This means that the image set of the correspondence about that point is strictly contained in some closed cube $K$, i.e. $\varphi(B_\varepsilon(x) \cap A) \subset$ $K$. For every sequence $\{x_n\}$ in $A$ converging to $x$ and every sequence $\{y_n\}$ in $B$ converging to $y$ such that for every $n \in N$, $y_n \in \varphi(x_n)$, one has $y \in \varphi(x)$. In other words, if the image set of the correspondence $\varphi$ approaches $y$ as one approaches $x$ from all possible directions in its domain, the correspondence is upper hemicontinuous at $x$ if $y$ belongs to its image set at that point. This means that the correspondence does not shrink at $x$. A correspondence is upper hemicontinuous if it is upper hemicontinuous at every point of its domain.

---

[13] See Arrow and Hahn (1971, p. 376).

The graph of an upper hemicontinuous correspondence is closed. The set of all possible pairs of points made up of a point of the domain of the correspondence and a point of its image set at that point of its domain, $G(\varphi) = \{(x,y) \in A \times B | y \in \varphi(x)\}$, makes up a set with no holes. That is, for an upper hemicontinuous correspondence $\varphi : A \to 2^B$, $\ell p\big(G(\varphi)\big) \subseteq G(\varphi)$. Conversely, if a correspondence that takes its values in a bounded set $-\varphi(A) \subset K-$ has a closed graph, it is upper hemicontinuous.[14]

A correspondence $\varphi$ from $A$ into $B$ is *lower hemicontinuous* at a point $x$ of $A$ if for every sequence $\{x_n\}$ in $A$ converging to $x$ and every $y \in \varphi(x)$ there is a sequence $\{y_n\}$ in $B$ converging to $y$ such that for every $n \in N$ one has $y_n \in \varphi(x_n)$. In other words, the correspondence $\varphi$ is lower hemicontinuous at $x$ if its image set approaches every point of its image set at $x$ as one approaches $x$ from every possible direction in its domain. This means that the correspondence does not spike at $x$. A correspondence is lower hemicontinuous if it is lower hemicontinuous over its domain.

A correspondence $\varphi$ from $A$ into $B$ is *continuous* at a point $x$ of $A$ if and only if it is both upper and lower hemicontinuous at that point. It is continuous if it is continuous at every point of its domain.

---

[14] See Green and Heller (1981, pp. 46, 47).

The correspondence $\varphi$ is *convex valued* at a point $x$ if and only if its image set at that point contains the segment joining any two points that belong to the image set. Formally, a correspondence $\varphi : A \to 2^B$ is convex valued at a point $x$ of $A$ if and only if $[y, z] \subseteq \varphi(x)$, for any pair of points $y, z \in \varphi(x)$. It is convex valued if it is convex valued over its domain.

The *non-negative orthant* of the Euclidean space of some finite dimension $l$ is denoted by $R^l_+$. It is the set of all points whose $l$ coordinates are either positive or zero. Formally, $R^l_+ = \{x \in R^l | x^j \geq 0, j = 1, \cdots, l\}$.[15] The *standard probability simplex* in that space, denoted by $P$, is the set of all points whose values add up to 1. Formally, $P = \{p \in R^l_+ | \sum_{j=1}^{l} p^j = 1\}$. Its dimension is one less than that of the Euclidean space that contains it, as all values are linearly related to add up to one. Thus, $\dim(P) = l - 1$.

Fixed points are defined for correspondences that go from a given set back into itself. A *fixed point* is a point of the domain of the correspondence that belongs to its image set at that point. That is, $x \in A$ is a fixed point of $\varphi : A \to 2^A$ if and only if $x \in \varphi(x)$. When either consumption or production are correspondences, the existence of equilibrium is proved through *Kakutani's fixed-point theorem,* which follows.

*Kakutani's fixed-point theorem*

If $A$ is a non-empty, compact, convex subset of a Euclidean space, and $\varphi$ is an upper hemicontinuous, convex valued correspondence from $A$ to $A$, then $\varphi$ has a fixed point.

---

[15] The non-positive orthant, denoted by $R^l_-$, is defined analogously.

# 3. Assumptions on the Excess Consumption Correspondence and the Exchange Value Correspondence

The traditional approach to proving the existence of a general equilibrium gives economic content to Kakutani's theorem so that it can be used to this end.[16]

The most general approach entails working with the so-called excess demand correspondence and postulating that it satisfies the assumptions of Kakutani's fixed point theorem.[17] The excess demand correspondence in that framework is the difference between net consumption and the total stocks of available resources. Net consumption is the stock of resources required to meet total consumption either directly or through production.[18] Therefore, this paper calls the excess demand correspondence the *excess consumption* correspondence instead. In a pure exchange framework without production, it is the amount by which the total planned consumption of each good exceeds or falls short of the total stocks available.

Excess consumption is indeed a correspondence. Although the total stocks of goods available at a given time are a vector or, equivalently, a point in a Euclidean space, each individual's planned consumption is a correspondence. The reason is that given a list of prices for the goods available, an individual might have a collection of alternative optimal consumption plans between which they are indifferent rather than just one.[19] As a result, total consumption is a correspondence since it is the sum of the consumption correspondences of all individuals. Each coordinate of these vectors represents one of the goods available at the time. The magnitude of each coordinate represents the quantity of the good in question.

---

[16] See Arrow and Debreu (1954), Debreu (1956, 1959, 1982), Arrow and Hahn (1971), Gale (1955), McKenzie (1959) and Uzawa (1962).

[17] See Debreu (1956; 1982, pp. 715-719).

[18] See Debreu (1959, p. 75).

Let $\zeta$ denote the excess consumption correspondence and $z$ a vector of excess consumption. Also, let $\xi_i$ denote the consumption correspondence of some individual $i$, $x^i$ a consumption vector for this individual and $h^i$ their initial endowment of goods. Assuming that there are $m$ individuals, $z = \sum_{i=1}^{m}(x^i - h^i)$. $x^i, h^i$ and $z$ are column vectors.

Let $p$ denote a list of what General Equilibrium Theory calls prices. This paper calls them exchange values instead for two reasons. Firstly, there is no money in General Equilibrium Theory, only a numeraire at the most – a good that officiates as the measure of the prices of all other goods. Secondly, what matters in General Equilibrium Theory are relative prices – the price of goods relative to one another – not absolute prices.[20] $p$ is a row vector made up of as many exchange values as there are goods available. Moreover, the exchange values in each list $p$ are normalised to take their values in the simplex $P$, so their sum always adds up to 1 in any such list.

All the above vectors but excess consumption vectors are non-negative. Their coordinates are either zero or positive but cannot be all zero since the coordinates of any list of exchange values add up to 1.[21] Assume there are $l$ different types of goods in total. Then, the Euclidean space in which all vectors take their values has dimension $l$.

---

[19] In this paper economic agents are individuals for convenience. However, the pure exchange framework can be extended to households.

[20] Excess consumption correspondences are always characterised as homogeneous of degree zero relative to the prices of goods. See Arrow and Debreu (1954), Debreu (1959, 1974, 1982), Arrow and Hahn (1971), Sonnenschein (1973), Mantel (1974), Shafer and Sonnenschein (1982), Dierker (1982) and Balasko (2016).

Formally,

$$x^i, h^i \in R_+^l$$

$$p \in P$$

$$P \subset R_+^l$$

$$z \in R^l$$

$$\xi_i : P \to 2^{R_+^l}$$

$$\zeta : P \to 2^{R^l}$$

Where

$$\zeta(p) = \sum_{i=1}^{m} \big(\xi_i(p) - h^i\big) = \{z = \sum_{i=1}^{m}(x^i - h^i) \,|\, x^i \in \xi_i(p), i = 1, \cdots, m\}$$

A general equilibrium is defined as a state where at least one of the vectors that make up the excess consumption correspondence at a given list of exchange values is non-positive. For that vector, the planned consumption of every good at that list of exchange values is no larger than the total stocks available.

Formally, the set of all possible equilibrium allocations of goods at a given $p_0$ can be defined as follows.

$$Z^*(p_0) = \{z \in \zeta(p_0) | z \leq 0\}$$

---

[21] This is in line with the standard understanding of prices and consumption. A negative price is actually a positive price offered for the disposal of the good. Also, in a pure exchange framework, consumption is non-negative because there is no production for others.

Or equivalently,

$$X^*(p_0) = \{\sum_{i=1}^{m} x^i \in \sum_{i=1}^{m} \xi_i(p_0) \mid \sum_{i=1}^{m} x^i \leq \sum_{i=1}^{m} h^i\}$$

Both sets will be empty if no general equilibrium allocation exists at the given list of exchange values $p_0$.

The excess consumption correspondence $\zeta$ is assumed to be upper hemicontinuous and convex valued over its domain. The problem, however, is that this correspondence does not go back into itself. Its domain is the exchange values of goods $P$. Given an individual's initial endowment, their consumption decisions hinge on the exchange values of goods for one another. The reason is that these values determine their budget set, that is, what they could afford at those exchange values. The image of the excess consumption correspondence at a list of exchange values $p_0$, in contrast, is the total consumption correspondence at that list of exchange values $\sum_{i=1}^{m} \xi_i(p_0)$ translated by the column vector of the total stocks of goods $\sum_{i=1}^{m} h^i$. It is the sum of the sets of alternative consumption bundles that would maximise each individual's satisfaction at a given list of exchange values shifted by the amount by which the total stocks of goods meet each sum of bundles. Both the domain and the image of the excess consumption correspondence at each point $p$ of its domain take their values in the same space $R^l$ but make up different subsets of $R^l$ altogether.

Let $Z$ denote the subset of all possible excess consumptions, $Z \subset R^l$. The correspondence to which Kakutani's theorem is applied to prove the existence of a general equilibrium is the Cartesian product of the excess consumption correspondence $\zeta$ and an exchange value correspondence that depends on excess consumption denoted by $\mu : Z \to 2^P$. Therefore, both the domain and codomain of the resulting correspondence are the Cartesian product of the simplex $P$ and subset $Z$. This derived correspondence is denoted by $\psi$.[22]

$$\psi : P \times Z \to 2^{P \times Z}$$

Specifically,

$$\psi(p, z) = \mu(z) \times \zeta(p)$$

As total consumption $\sum_{i=1}^{m} \xi_i$ is a correspondence, the difference between total consumption at any given list of exchange values $p$ and the total stocks of goods $\sum_{i=1}^{m}(\xi_i(p) - h^i)$ is a set of vectors $\{z \in \zeta(p)\}$. Each vector $z$ is the gap between one total consumption $\sum_{i=1}^{m} x_i \in \sum_{i=1}^{m} \xi_i(p)$ which at the given list of exchange values $p$ is optimal for each consumer $i$ and the total stocks of goods $\sum_{i=1}^{m} h^i$.

The excess consumption correspondence $\zeta$ being upper hemicontinuous means the following. The excess consumption correspondence is closed and bounded, and hence compact.[23] In other words, there are a maximum and a minimum value for the gaps between the total consumption and the total stocks of each good. This means that for each $j = 1, \cdots, l$ there are two finite real numbers $z_1^j$ and $z_2^j$ such that they are coordinates of two gaps $z_1 \in \zeta(p_1), z_2 \in \zeta(p_2)$ for some $p_1, p_2 \in P$ and $z_1^j \leq z^j \leq z_2^j$ for every $z \in \zeta(p)$ for all $p \in P$. Additionally, suppose a sequence of gaps $\{z_n\}$ such that for each $n \in N$, $z_n \in \zeta(p_n)$, tends to a certain point $z$ as the list of exchange values for those goods $\{p_n\}$

---

[22] The Cartesian product of two correspondences is defined analogously to the Cartesian product of two sets. It is the set of all possible ordered pairs of points but across the image set of each correspondence at given points in their respective domains. Formally,

$\mu(z') \times \zeta(p') = \{(p, z) \in P \times Z | p \in \mu(z') \text{ and } z \in \zeta(p') \text{ for some } z' \in Z \text{ and some } p' \in P\}$.

[23] See Debreu (1982, pp. 698, 699) and Green and Heller (1981, pp. 46, 47).

approaches some given list $p$ as $n \to +\infty$. Then, that point is a gap between total consumption at that list of exchange values and the total stocks of goods, i.e. $z \in \zeta(p)$.[24]

The excess consumption correspondence being convex valued, in turn, has the following meaning. For any two gaps between total consumption at a given list of exchange values and the total stocks of goods $z_1, z_2 \in \zeta(p)$, every point in between is also a gap at that list of exchange values, i.e. $[z_1, z_2] \subseteq \zeta(p)$. Therefore, each belongs to the excess consumption correspondence at that list of exchange values.

Now the exchange value correspondence $\mu$ must be defined and proved to satisfy the assumptions of Kakutani's theorem itself. Also, the domains and codomains of the excess consumption correspondence $\zeta$ and the exchange value correspondence $\mu$, which are the simplex $P$ and the set of all possible excess consumption vectors $Z$ respectively, must be proven to be compact and convex.

First, let us look at the simplex $P$. The simplex $P$ is defined as the set of all possible lists of exchange values such that each exchange value on any such list is non-negative and their sum adds up to 1. The simplex $P$ allows representing all lists of exchange values as a compact convex set since it is closed, bounded and convex by definition. It is also trivially non-empty.

**Proof of boundedness**

Since $p > 0$ for all $p \in P$, $p^j \geq 0, j = 1, \cdots, l$. Also, $\sum_{j=1}^{l} p^j = 1$ for all $p \in P$. Therefore, $0 \leq p^j \leq 1, j = 1, \cdots, l$ for all $p \in P$.

---

[24] The existence of such convergent sequences is guaranteed by the fact that every infinite sequence in a bounded space has a convergent subsequence. The excess consumption correspondence $\zeta$ has been assumed upper hemicontinuous, and hence bounded. Next it will be shown that the simplex $P$ is bounded too.

**Proof of closedness**

Let $\{p_n\}$ be a sequence converging to some $p$ such that for all $n \in N$, $p_n \in P$. Such a sequence exists because $P$ is bounded. Assume $p \notin P$. Therefore, for $\varepsilon'' > 0$ small enough there is an open ball $B_{\varepsilon''}(p)$ of radius $\varepsilon''$ about $p$ in which either $p_k^j < 0$ for some $j = 1, \cdots, l$ or $\sum_{j=1}^{l} p_k^j \neq 1$ for all $p_k \in B_{\varepsilon''}(p)$. By the definition of the limit of a convergent sequence, for any number $\varepsilon > 0$ there exists a positive integer $n'$ such that for all $n > n'$, $|p_n^j - p^j| < \varepsilon$ for all coordinates $j = 1, \cdots, l$. Therefore, there exists some $n''(\varepsilon'') \in N$ such that for all $n > n''$ either $p_n^j < 0$ for some $j = 1, \cdots, l$ or $\sum_{j=1}^{l} p_n^j \neq 1$. However, this contradicts the assumption that $p_n \in P$ for all $n \in N$. Therefore, $p \in P$. Since $p$ is arbitrary, $\ell p(P) \subseteq P$ and $P$ is closed.

$P$ is also a convex set since it contains the segment joining any two points in it. The reason is that every convex combination of points whose coordinates are non-negative and add up to 1 is itself non-negative and made up of elements which add up to 1.

**Proof of convexity**

Let $p_1$ and $p_2$ be two arbitrary points in $P$ and let $0 \leq \alpha \leq 1$. Any convex combination of $p_1$ and $p_2$ can be written as $p = \alpha p_1 + (1 - \alpha) p_2$ for some $\alpha \in [0,1]$. Since $p_1, p_2 > 0$ by hypothesis, $\alpha \cdot p_1^j + (1 - \alpha) \cdot p_2^j \geq 0$ for $j = 1, \cdots, l$. Additionally, $\sum_{j=1}^{l}(\alpha \cdot p_1^j + (1 - \alpha) \cdot p_2^j) = \alpha \cdot \sum_{j=1}^{l} p_1^j + (1 - \alpha) \cdot \sum_{j=1}^{l} p_2^j$. Since $\sum_{j=1}^{l} p_1^j = \sum_{j=1}^{l} p_2^j = 1$ by hypothesis, $\alpha \cdot \sum_{j=1}^{l} p_1^j + (1 - \alpha) \cdot \sum_{j=1}^{l} p_2^j = 1 \cdot (\alpha + 1 - \alpha) = 1$.

The simplex $P$ defines the subset of all possible lists of exchange values. This set is the domain of the excess consumption correspondence $\zeta$ and the codomain of the exchange value correspondence $\mu$. The latter is still to be defined and proven to be upper hemicontinuous and convex valued. This correspondence is defined as the set of lists of exchange values in the simplex that maximise the value of any given excess consumption.[25]

$$\mu(z) = \left\{ p \in P \,\middle|\, p \cdot z = \max_{p \in P} p \cdot z \right\}$$

Given any arbitrary $z \in Z$, $\mu(z) \neq \emptyset$ since $P$ is compact. The exchange value correspondence $\mu$ is non-empty for any given vector $z$ of excess consumption. To prove that it is upper hemicontinuous, the following lemma from Debreu (1982, p. 701) can be used.

*Lemma 1*

Let $A$ and $B$ be subsets of Euclidean spaces, $f : A \times B \to R$ be a real-valued function and $\varphi : B \to 2^A$ and $\mu : B \to 2^A$ be correspondences such that $\mu(b) = \left\{ a \in \varphi(b) \,\middle|\, f(a,b) = \max_{c \in \varphi(b)} f(c,b) \right\}$, $b \in B$. If the function $f$ and the correspondence $\varphi$ are continuous, then the correspondence $\mu$ is upper hemicontinuous.

The lemma means that if both the function to be maximised and the correspondence from where $\mu$ takes its set of values are continuous, then the correspondence $\mu$ is upper hemicontinuous. Let us define both.

$$f : P \times Z \to R$$

Specifically,

$$f(p, z) = p \cdot z$$

---

[25] Debreu (1982, p. 708) justifies the exchange value correspondence on the grounds that the market increases the exchange values of goods with positive excess consumption and decreases those of goods with negative excess consumption. As a result, the total value of the excess consumption correspondence is increased. See also Arrow and Debreu (1954, p. 271).

Additionally,

$$\varphi : Z \to 2^P$$

Specifically,

$$\varphi(z) = P$$

The function $f = p \cdot z = \sum_{j=1}^{l} p^j z^j$ is continuous since addition and multiplication are continuous operations. The correspondence $\varphi$ is constant and, hence, trivially continuous. Whatever excess consumption is, it is possible to choose any list of exchange values from the simplex $P$. Thus, the conditions of *Lemma 1* are satisfied and the exchange value correspondence $\mu$ is upper hemicontinuous. Take a sequence of excess consumptions $\{z_n\}$ in $Z$ converging to $z$ as $n \to +\infty$ and any sequence of lists of exchange values $\{p_n\}$ in $P$ with the following property.[26] For every $n$, the list of exchange values $p_n$ maximises the value of the excess consumption $z_n$. That is, $p_n \cdot z_n = \max_{p \in P} p \cdot z_n$. By definition, $p_n \in \mu(z_n)$ for every $n$. Then the resulting sequence of exchange values $\{p_n\}$ has some limit $p$, which both belongs to $P$ and maximises the value of the excess consumption vector $z$. Hence, $p$ belongs to the image set of $\mu(z)$.

Moreover, the correspondence $\mu$ is convex valued because for any given excess consumption $z'$, it is the set of maximisers of a linear function - $f(p) = p \cdot z'$ - on a convex set - $P$. Thus, if two different lists of exchange values maximise the value of a given excess consumption, so does every list in between.

**Proof**

Let $p_1, p_2 \in P$ be two lists of exchange values such that for some excess consumption vector $z'$, $p_1 \cdot z' = p_2 \cdot z' = \max_{p \in P} p \cdot z'$, $p_1 \neq p_2$. Also, let $0 \leq \alpha \leq 1$.

---

[26] The boundedness of set $Z$ is discussed further down this section.

$(\alpha p_1 + (1-\alpha)p_2) \cdot z' = \alpha p_1 \cdot z' + (1-\alpha)p_2 \cdot z'$ Since $p_1 \cdot z' = p_2 \cdot z'$ by hypothesis, $\alpha p_1 \cdot z' + (1-\alpha)p_2 \cdot z' = \alpha p_1 \cdot z' + (1-\alpha)p_1 \cdot z' = (\alpha + 1 - \alpha)p_1 \cdot z' = p_1 \cdot z'$. Again, by hypothesis $p_1 \cdot z' = p_2 \cdot z' = \max_{p \in P} p \cdot z'$.

As $P$ is a non-empty compact set, the assumption that the excess consumption correspondence $\zeta$ is upper hemicontinuous on $P$ means that there is a non-empty bounded subset $Z \subset R^l$ such that $\zeta(p) \subset Z$ for all $p \in P$. We can take the subset $Z$ to be compact and convex.[27]

Thus, the correspondence $\psi(p, z) = \mu(z) \times \zeta(p)$ satisfies all the assumptions of *Kakutani's fixed-point theorem* and has at least one fixed point. This means that there is a point $(p^*, z^*) \in P \times Z$ such that $p^* \in \mu(z^*)$ and $z^* \in \zeta(p^*)$.

However, one more assumption must be made to ensure that the fixed point is a general equilibrium.

*Walras's law*

$$p \cdot \zeta(p) \leq 0$$

In a private ownership economy, the exchange value of total consumption at the going list of exchange values can never exceed the exchange value of the total stocks of goods available. The latter includes both non-produced and produced stocks at those exchange values.

Walras's law implies that the excess consumption vector at any fixed point is non-positive. That is, $z^* \leq 0$. Hence, it is an equilibrium excess consumption from an economic point of view.

---

[27] See Debreu (1982, p. 718).

**Proof**

$p \cdot z^* \leq p^* \cdot z^*$ since $p^* \in \mu(z^*)$. Moreover, $p^* \cdot z^* \leq 0$ since $\zeta(p)$ obeys Walras's Law. Therefore, $p \cdot z^* \leq 0$ for all $p \in P$. Since $\{0\} \notin P$, $p \cdot z^* \leq 0$ for all $p \in P$ means that $z^* \in R^l_-$.

Take a set of vectors that normalises all possible non-negative vectors bar vector $\vec{0}$. The product of each vector in this set and some other vector can only be non-positive if the other vector is non-positive itself. Therefore, each coordinate of the excess consumption vector $z^*$ at a fixed point is either zero or negative.[28]

## 4. Assumptions on Individuals

The next step is to work out conditions on individuals' preferences and the constraints they face so that the *excess consumption correspondence* is indeed *upper hemicontinuous* and *convex valued* and satisfies *Walras's Law*. In a pure exchange framework, it is enough for the consumption correspondence of individuals to be upper hemicontinuous and convex valued since the total stocks of goods are given and hence constant. Additionally, the exchange value of every sum of individual consumption bundles across the consumption correspondences of individuals at an arbitrary list of exchange values should never exceed the exchange value of the given total stocks of goods. These conditions should also ensure that all possible equilibrium lists of exchange values are non-negative and different from zero. As the below conditions are sufficient rather than necessary, they are stronger than simply assuming that the excess consumption correspondence and the lists of exchange values have the desired properties.[29]

---

[28] Debreu (1956; 1982, pp. 717-719) proves a more general theorem. The exchange values of goods are normalised to take their values in the unit sphere. Take any non-degenerate pointed closed convex cone with vertex 0. That is any cone from the origin whose vertex has an internal angle greater than zero degrees and smaller than 180 degrees. The lists of exchange values could be any point in the section of the unit sphere that lies within that cone. He proves that there is always a list of exchange values in that section of the unit sphere such that the following happens. At least one of the vectors of the excess consumption correspondence lies within the polar of the cone. Therefore, his proof allows for the possibility that some of the equilibrium exchange values could be negative. This paper has argued that a negative exchange value signals the demand for the disposal of a good at a positive exchange value. As a result, it is enough to restrict the cone to the positive orthant of the goods space.

Firstly, a consumption set $X_i$ is defined for each individual $i$, which represents all the possible consumptions for that individual, including no consumption. This set is assumed closed, convex and bounded from below.

The consumption set of an individual - $X_i$ - being closed guarantees the following. If a sequence of possible consumptions for an individual $\{x_n\}$ tends to some bundle of goods $x$ as $n \to +\infty$, then bundle $x$ is also a possible consumption for them, i.e. $x \in X_i$.[30] $X_i$ being convex means that the set of possible consumptions for an individual includes every bundle of goods between any two bundles in the set. Formally, if $x_1, x_2 \in X_i$, then $[x_1, x_2] \subseteq X_i$. It also implies that goods are perfectly divisible. An individual's consumption set being bounded from below ensures that there is a set of consumptions below which the individual's consumption cannot fall. This assumption fits in with the fact that consumption is non-negative.[31]

---

[29] See Debreu (1982, p. 722).

[30] Put differently, $\ell p(X_i) \subseteq X_i$.

[31] General Equilibrium Theory allows negative consumption for the sake of generality. However, this detracts from the theory rather than enhances it. Firstly, it clashes with the commonsense understanding of consumption as the depletion of positive quantities of goods. Secondly, it conflates consumption with production. At any rate, in a framework of pure exchange without production, consumption can only be non-negative.

Secondly, it is assumed that everyone maximises their own preferences. It is further assumed that individuals' preferences are continuous. For any consumption bundle in an individual's consumption set, the set of consumption bundles they desire as much or more is closed. Likewise, the set of consumption bundles they desire as much or less is also closed.[32]

Formally, for every $i$ the set $\{(x, x') \in X_i \times X_i | x \precsim x'\}$ is closed.

This assumption means the following. Take every sequence of consumption bundles $\{x_n\}$ in the consumption set $X_i$ of some individual $i$ tending to some consumption bundle $x$ such that they never prefer bundle $x_n$ to some other bundle $x'$ for any $n \in N$. Then this individual will not prefer consumption bundle $x$ to $x'$ either. Likewise, for any convergent sequence such that this individual does not prefer bundle $x'$ to any bundle in the sequence, they will not prefer it to the bundle that the sequence tends to either.

Formally, for every $\{x_n\} \subset X_i$ tending to $x \in X_i$ as $n \to +\infty$ and some $x' \in X_i$ such that for every $n \in N$, $x_n \precsim x'$, $x \precsim x'$.

Likewise, for every $\{x_n\} \subset X_i$ tending to $x \in X_i$ as $n \to +\infty$ and some $x' \in X_i$ such that for every $n \in N$, $x_n \succsim x'$, $x \succsim x'$.

The assumptions that the consumption set $X_i$ of individual $i$ is convex – hence connected – and their preference relation $\succsim_i$ closed allows representing their preferences through a continuous function $u_i: X_i \to R$, called a utility function. That means there are no jumps in the individual's satisfaction levels as they move over their set of possible consumptions. The utility function has the property that a consumption bundle represents no less utility than some other only if the individual does not prefer the other. That is, $u_i(x') \geq u_i(x)$ if and only if $x' \succcurlyeq_i x$. The function $u_i$ is uniquely determined up to a strictly increasing monotonic transformation. If $u_i$ is a utility function for individual $i$, so is $f(u_i)$ provided that $f$ is a real-valued strictly increasing function of $u_i$. This defines the utility functions of individuals as ordinal as opposed to cardinal. Any function that keeps the order of preferences for individual $i$ is a utility function for that individual. Moreover, if $f$ is continuous, so will be $f(u_i)$ on $X_i$.[33]

---

[32] See Arrow and Debreu (1954, p. 269), Debreu (1959, p. 56; 1982, pp. 705, 711), Arrow and Hahn (1971, p. 78), Barten and Böhm (1982, p. 386) and Balasko (2016, p. 48).

Given the above definition of the utility function for individual $i$, the first part of *Lemma 1* is satisfied. Individuals maximise their preferences, which means maximising a continuous utility function. The next step is to define sufficient conditions so that each individual chooses their preferred consumption bundles on a continuous correspondence.

In a private ownership economy, everybody must support themselves, so everyone must stick to a budget. The budget set of an individual is defined as all the alternative combinations of goods within their consumption set that they could afford. Given the individual's goods endowment, it depends on the list of the exchange values of goods at that time.[34] To satisfy the second part of *Lemma 1*, sufficient conditions must be worked out for each individual's budget set to be continuous over the range of all possible lists of exchange values.

---

[33] See Debreu (1959, pp. 56-59), Arrow and Hahn (1971, pp. 82-87) and Barten and Böhm (1982, pp. 388-390).

[34] See Debreu (1959, pp. 79, 81, 86; 1982, pp. 704, 706, 708, 710, 715, 716) and Arrow and Hahn (1971, pp. 77, 80, 107).

Let $\beta_i$ denote the budget set of some individual $i$. $\beta_i: P \to X_i$ is defined as $\beta_i(p) = \{x^i \in X_i | p \cdot x^i \leq p \cdot h^i\}$.

General Equilibrium Theory understands that goods with no exchange value are freely available. They are what the theory calls free goods.[35] That understanding guarantees that the budget set is upper hemicontinuous at every list of exchange values where it is not empty, provided that the set on which it is defined is compact.[36] At certain lists of exchange values, however, there might be no possible consumption that the individual could afford. In other words, there might be some $p' \in P$ such that $\min_{x^i \in X_i} p' \cdot x^i > p' \cdot h^i$. Then, $\beta_i(p') = \emptyset$. Moreover, the consumption set of individual $i$ - $X_i$ - is assumed to be closed and bounded from below. To be compact, it must also be bounded from above.

The total stocks of goods are finite and everybody's consumption set is bounded from below. Both features together mean that everybody's *feasible consumption* is bounded from above.[37] The reason is that nobody can increase their consumption indefinitely on the back of unbounded negative consumption by other individuals. Thus, it is natural to focus on a compact subset of each individual's consumption set that contains their set of feasible consumptions in its interior. Let $X'_i$ denote this set for some individual $i$. Before the consumption of the individual reaches any section of the upper boundary, the bounding goods will have a positive exchange value due to scarcity. For any such goods, the individual's budget set itself creates an upper bound on the maximum amounts of these goods that the individual could afford.

---

[35] See Arrow and Debreu (1954, pp. 271, 272, 288), Debreu (1959, p. 33) and Arrow and Hahn (1971, p. 9).

[36] See Arrow and Debreu (1954, pp. 277, 278), Debreu (1959, pp. 63, 64; 1982, p. 707) and Arrow and Hahn (1971, pp. 81, 109).

[37] See Debreu (1959, p. 77) and Arrow and Hahn (1971, p. 89).

The graph of the budget set of an individual $i$ is $\{(p, x^i) \in P \times X_i | p \cdot x^i \leq p \cdot h^i\}$. Even if goods are freely available when they lose all exchange value, the graph of the budget set on the compact subset $X_i'$ is closed and bounded at every list of exchange values where it is not empty. Therefore, the budget set $\beta_i$ is upper hemicontinuous on $X_i'$.

**Proof**

The consumption set $X_i$ is closed and the dot products $p \cdot x^i = \sum_{j=1}^{l} p^j x^{ji} : P \times X_i \to R_+$ and $p \cdot h^i = \sum_{j=1}^{l} p^j h^{ji} : P \to R_+$ are continuous over their domains since addition and multiplication are continuous operations. Moreover, $\beta_i$ is defined by the weak inequality $p \cdot x^i \leq p \cdot h^i$ on the closed set $X_i$ for every $p \in P$ and subset $X_i' \subset X_i$, is not only closed but also compact. Debreu (1982, p. 706) shows that the function $f(p) = \min_{x^i \in X_i} p \cdot x^i$ is continuous provided that $X_i$ is a non-empty, compact subset of $R^l$, which $X_i'$ is. As $p \cdot h^i$ is also continuous over $P$, the set $P' = \left\{ p \in P \middle| p \cdot h^i \geq \min_{x^i \in X_i'} p \cdot x^i \right\}$ is closed. Then the graph of the budget set is closed on $P' \times X_i'$. Finally, since $X_i'$ is compact, the graph of the budget set is bounded in the neighbourhood of every $p \in P'$. Therefore, the budget set is upper hemicontinuous on $X_i'$.

However, understanding goods with no exchange value as free goods jeopardises the lower hemicontinuity of the budget set even though the consumption set $X_i$ of each individual is convex.[38]

The exchange value of the individual's goods endowment might be equal to that of their cheapest possible consumption at the going list of exchange values. Formally, $p \cdot h^i = \min_{x^i \in X_i} p \cdot x^i$. In that case, their budget set might not be lower hemicontinuous at that list of exchange values.

---

[38] See Debreu (1959, p. 63) and Arrow and Hahn (1971, p. 80).

Suppose the exchange value of a good drops to zero as the list of the exchange values of goods changes within the set $P'$ to reach $p$. That is, for some sequence $\{p_n\} \subset P'$ tending to $p \in P'$ as $n \to +\infty$ and some good $j = 1, \cdots, l$, $p_n^j > 0$ while $p^j = 0$. Assume that the individual has the minimum possible consumption of all other goods combined but some extra quantities of good $j$. Then they could only swap good $j$ for some more quantities of the other goods. However, when its exchange value hits zero, the individual can no longer demand any goods with it but can afford unlimited quantities of this good because it has become a free good. Take a bundle $x^i \in X_i'$ which is equal to their initial goods endowment except that it has more quantities of good $j$. That is, $x^i > h^i$ because $x^{ji} > h^{ji}$ while $x^{ki} = h^{ki}$ for each good $k \neq j$. For no bundle such as $x^i$ is there a sequence of bundles $\{x_n^i\}$ that tends to bundle $x^i$ as $n \to +\infty$ such that for each $n \in N$, $x_n^i \in \beta_i(p_n)$. In other words, from the sequence of budget sets that the sequence of exchange values tending to $p$ gives rise to, no sequence of bundles can be extracted that tends to $x^i$.

**Proof**

Assume there were one such sequence $\{x_n^i\}$ that tends to $x^i \in \beta_i(p)$ such that $x_n^i \in \beta_i(p_n)$ for some sequence $\{p_n\} \subset P'$ tending to $p \in P'$ as $n \to +\infty$. As $P'$ and $G(\beta_i)$ on $P' \times X_i'$ are compact, the existence of both convergent sequences is guaranteed. For $n$ large enough, say $n > n'$, the sequence $\{x_n^i\}$ would be interior to $\beta_i(p)$ since $x^{ji} > h^{ji}$ while $x^{ki} = h^{ki}$ for each good $k \neq j$. However, this implies that $p \cdot x_n^i < p \cdot x^i$ for $n > n'$. Therefore, the assumption that $p \cdot h^i = \min_{x^i \in X_i} p \cdot x^i$ is contradicted since $p \cdot h^i = p \cdot x^i$ by hypothesis.

Debreu (1982, p. 707) provides sufficient conditions for the budget set of individuals to be lower hemicontinuous at some list of exchange values $p_0$ in the following lemma. As it is upper hemicontinuous when restricted to a compact subset of the individual's consumption set $X_i$, the lemma guarantees that the budget set is continuous at $p_0$.

*Lemma 3*

If $X_i$ is non-empty, compact and convex, and $p_0 \cdot h^i > \min_{x^i \in X_i} p_0 \cdot x^i$, then the correspondence $\beta_i$ is continuous at $p_0$.

If *Lemma 3* holds at every list of exchange values, the second half of *Lemma 1* is satisfied. Thus, *Lemma 3* and the assumptions that preferences are continuous and everyone's consumption set is connected provide sufficient conditions for the consumption correspondence of individuals to be upper hemicontinuous. For some individual $i$, their consumption correspondence is the set of alternative consumption bundles that would maximise their utility from those they can afford at the given list of exchange values $p$.

Formally, $\xi_i : P \to 2^{X_i}$ and is defined as

$$\xi_i(p) = \left\{ x \in \beta_i(p) \middle| u(x) = \max_{x' \in \beta_i(p)} u(x') \right\}$$

As the total consumption correspondence is the sum of the consumption correspondences of individuals, the total consumption correspondence will also be upper hemicontinuous. So will the excess consumption correspondence since it is merely the total consumption correspondence translated by the vector of total goods endowments.

Next, sufficient conditions must be worked out to ensure that *Lemma 3* is satisfied at all lists of exchange values. Debreu (1982, p. 705) provides the strongest and simplest condition. Everyone's initial goods endowment is strictly larger than one of their possible consumptions.

Formally, for every $i$ there is $x_0^i \in X_i$ such that $x_0^i \ll h^i$.

As each individual's goods endowment is strictly larger than at least one possible consumption, it is larger in each coordinate. This means not only that each individual could potentially be self-sufficient but that their endowment has some quantities of every good available at the time. As a result, the exchange value of their endowment will always be positive. Since $p > 0$, $x_0^i \ll h^i$ means that $p \cdot x_0^i < p \cdot h^i$ at all $p \in P$. Therefore, the excess consumption correspondence is upper hemicontinuous by *Lemma 1*. Moreover, $P' = P$, so each individual's budget set $\beta_i, i = 1, \cdots, m$ is non-empty, and hence defined, over the entire exchange value simplex $P$. That is, for every $p \in P$ and every , $i = 1, \cdots, m$, $\beta_i(p) \neq \emptyset$.[39]

The budget set of individuals also plays a second role. It guarantees the fulfilment of *Walras's Law*.

Each individual faces the following constraint.

$$p \cdot x^i \leq p \cdot h^i$$

Adding up across all individuals yields

$$\sum_{i=1}^{m} p \cdot x^i \leq \sum_{i=1}^{m} p \cdot h^i$$

Therefore,

$$p \cdot \sum_{i=1}^{m}(x^i - h^i) \leq 0$$

Since by definition,

---

[39] The existence of equilibrium is also proved under weaker assumptions, such as everyone being able to produce at least one desired good (Arrow & Debreu, 1954, pp. 280, 281) or being resource related to one another (Arrow & Hahn, 1971, pp. 117-119) or the economy being irreducible (McKenzie, 1959, pp. 55, 58, 59). However, the goal of these alternative assumptions is always to make the budget set of every individual continuous at every feasible list of exchange values.

$$\sum_{i=1}^{m}(x^i - h^i) = z$$

Then,

$$p \cdot z \leq 0$$

As the above inequality holds for all consumptions that each individual could afford at the list of exchange values $p$ - $\{x^i \in \beta_i(p)\}$ - it also holds for their preferred consumptions $\{x^i \in \xi_i(p)\}$.

Therefore,

$$p \cdot \zeta(p) \leq 0$$

The next step is to provide sufficient conditions for the consumption correspondence of individuals to be convex valued. To this end, the semi-strict convexity of their preference relation is assumed. Take two consumption bundles in an individual's consumption set such that they strictly prefer one to the other. Then they will also strictly prefer any convex combination of these two bundles to the least preferred of both.[40]

Formally, if $x$ and $x'$ are two points of $X_i$ such that $x \prec_i x'$, and $\alpha$ is a real number in $]0,1]$, then $x \prec_i (1-\alpha)x + \alpha x'$.

Debreu (1959, p. 60) and Arrow and Hahn (1971, p. 79) show that if the preference relation is semi-strictly convex and closed, it is also convex. Everybody either prefers any convex combination of two consumption bundles between which they are indifferent or is indifferent between them.

---

[40] See Arrow and Debreu (1954, p. 269), Debreu (1959, p. 84; 1982, pp. 705, 711) and Arrow and Hahn (1971, p. 78).

Formally, if $x$ and $x'$ are two points of $X_i$ such that $x \sim_i x'$, and $\alpha$ is a real number in [0,1], then

$x \preccurlyeq_i (1-\alpha)x + \alpha x'$

The convexity of the preference relation of individuals is a sufficient condition to guarantee that their consumption correspondences are convex valued. The consumption correspondence of an individual $i$ at any list of exchange values $p \in P$ is the intersection of two sets: the individual's budget set at that list and their preference relation. If $x'$ is a utility-maximising bundle for the individual within their budget set $\beta_i(p')$ for some $p' \in P$, then $\xi_i(p') = \beta_i(p') \cap \{x \in X_i | x' \precsim x\}$. The budget set $\beta_i$ is convex at every list of exchange values $p$ where it is not empty. The intersection of two convex sets, in turn, is convex.

**Proof that the budget set is convex**

Take two bundles $x_1, x_2 \in \beta_i(p')$ for some $p' \in P$ and let $0 \leq \alpha \leq 1$.

Firstly, $\alpha x_1 + (1-\alpha)x_2 \in X_i$ since $X_i$ is convex by hypothesis.

Secondly, $p' \cdot (\alpha x_1 + (1-\alpha)x_2) = \alpha p' \cdot x_1 + (1-\alpha)p' \cdot x_2$.

$\alpha p' \cdot x_1 \leq \alpha p' \cdot h^i$ since $p' \cdot x_1 \leq p' \cdot h^i$ and $\alpha \geq 0$ by hypothesis.

Likewise, $(1-\alpha)p' \cdot x_2 \leq (1-\alpha)p' \cdot h^i$ since $p' \cdot x_2 \leq p' \cdot h^i$ and $\alpha \leq 1$ by hypothesis, so $(1-\alpha) \geq 0$.

Then, $\alpha p' \cdot x_1 + (1-\alpha)p' \cdot x_2 \leq \alpha p' \cdot h^i + (1-\alpha)p' \cdot h^i$

Since $\alpha p' \cdot h^i + (1-\alpha)p' \cdot h^i = (\alpha + 1 - \alpha)p' \cdot h^i = p' \cdot h^i$, we have $p' \cdot (\alpha x_1 + (1-\alpha)x_2) \leq p' \cdot h^i$.

Therefore, $\alpha x_1 + (1-\alpha)x_2 \in \beta_i(p')$.

Two final assumptions are needed to guarantee that the equilibrium exchange values of goods are non-negative and not all zero so that they can be normalised to take their values in the simplex $P$.

These are the assumptions of local non-satiation and that any unused goods can be disposed of for free.

First, let us define the set of feasible consumptions of an individual and denote it $\hat{X}_i$. These are the consumption bundles in their consumption set that, added to the sum of the consumption bundles for all other individuals, never exceed the total stocks of goods available at the time. This process is repeated over all alternative ways that the existing stocks of goods could be allocated among the other individuals while ensuring that they are possible consumptions for each individual.[41]

In symbols, for every $i$

$$\hat{X}_i = \left\{ x^i \in X_i \,\middle|\, \text{there exists } x^h \in X_h \text{ for each } h \neq i \text{ such that } x^i + \sum_{\substack{h=1 \\ h \neq i}}^{m} x^h \leq \sum_{h=1}^{m} h^h \right\}$$

The first of the last two assumptions states that nobody can be satiated within their set of feasible consumptions.[42]

Formally, for every $i$, there is no satiation consumption in $\hat{X}_i$.

As the consumption set of every individual is bounded from below, this notion of insatiability is weaker than claiming that individuals can never be satiated. For example, if consumption is non-negative as this paper argues, the assumption means that no one can be satiated with the total stocks of goods available at the time.

---

[41] This formulation of the concept of an individual's feasible consumption set is a simplification of the original formulation (Arrow & Debreu, 1954, p. 276; Arrow & Hahn, 1971, pp. 88, 89; Debreu, 1959, p. 76; 1982, p. 705). In its original form, the concept also takes into account feasible productions. However, there is no production in a pure exchange framework.

[42] See Arrow and Debreu (1954, p. 269), Debreu (1959, p. 83; 1982, pp. 705, 711) and Arrow and Hahn (1971, p. 78).

The second assumption states that all individuals as a whole can dispose of any quantities of any goods for free. Let $Y_i$ be the goods disposal capacity of individual $i$, that is, all the different combinations of goods that this individual could dispose of. It is assumed that this set is contained in the goods space but is non-positive. The reason is that whatever is thrown out reduces the total stocks of goods available and, therefore, cannot be used for consumption. It is further assumed that everyone could keep everything they have if they wanted to. Moreover, some individuals might be unable to dispose of waste.[43]

Formally, for every $i$

$$Y_i \subset R_-^l$$

$$0 \in Y_i$$

Additionally,

$$\sum_{i=1}^{m} Y_i = R_-^l$$

Excess consumption vectors are now defined as $z = \sum_{i=1}^{m} x^i - \sum_{i=1}^{m} y^i - \sum_{i=1}^{m} h^i$ for some $x^i \in X_i$ and some $y^i \in Y_i$, $i = 1, \cdots, m$. Let $\eta_i$ denote the goods disposal choice of individual $i$. Given a list of exchange values $p \in P$, individual $i$ maximises their return on goods disposal in order to push out their budget constraint. Thus, $\eta_i$ is a correspondence $\eta_i : P \to 2^{Y_i}$. Specifically, $\eta_i(p) =$ $\left\{ y \in Y_i \middle| p \cdot y = \max_{y' \in Y_i} p \cdot y' \right\}$. The excess consumption correspondence is now $\zeta(p) = \sum_{i=1}^{m} \xi_i(p) - \sum_{i=1}^{m} \eta_i(p) - \sum_{i=1}^{m} h_i$.

---

[43] These assumptions are a simplification of the traditional assumptions on the production set to ensure the existence of equilibrium. See, for example, Debreu (1959, pp. 42, 84; 1982, p. 711) and Arrow and Hahn (1971, pp. 20, 70, 71). It is used here solely to account for free disposal.

The set of total possible goods disposal is closed, convex and bounded from above. The reason is that $\sum_{i=1}^{m} Y_i = R_{-}^{l}$. Moreover, the set of possible goods disposal which is also feasible given the total stocks of goods available is also bounded from below. Even if nobody consumed anything, there would only be the total stocks of goods to dispose of. Therefore, attention can be restricted to a compact, convex subset of $\sum_{i=1}^{m} Y_i$ which owns the element $\{0\}$ where nobody throws anything out. Let us denote this set $Y'$. The correspondence $\varphi : P \to 2^{R_{-}^{l}}$ such that $\varphi(p) = Y'$ for all $p \in P$ is constant and therefore trivially lower hemicontinuous. Since $Y'$ is compact, $\varphi$ is also upper hemicontinuous and hence continuous over its domain. $p \cdot \sum_{i=1}^{m} y^i$, in turn, is continuous over its domain since addition and multiplication are continuous operations. Therefore, the conditions of *Lemma 1* are satisfied and $\sum_{i=1}^{m} \eta_i(p)$ is an upper hemicontinuous correspondence. It is also convex valued. For any list of exchange values $p \in P$, $\sum_{i=1}^{m} \eta_i(p)$ is the outcome of maximising the linear function $p \cdot \sum_{i=1}^{m} y^i$ on the convex set $Y'$.

Let us now look at the role that the assumptions of local non-satiation and free disposal play. The assumption of local non-satiation rules out the possibility that an equilibrium list of exchange values might be zero. That is, $p^* \neq 0$. If, in contrast, the equilibrium list of exchange values were zero, nobody would be constrained in their consumption and everyone would choose unfeasible consumption bundles. The assumption of free disposal, in turn, ensures that all equilibrium lists of exchange values are non-negative, i.e. $p^* \geq 0$. The reason is that nobody would give anything in exchange for getting rid of unused quantities of goods when they can be freely disposed of.

Additionally, $\max_{y^i \in Y_i} p \cdot y^i = 0$ for all $p \in P$, $i = 1, \cdots, m$, since $p > 0$, $Y_i \subset R_-^l$ and $\{0\} \in Y_i$. Therefore, for $\{y^i \in \eta_i(p)\}$, $y^{ji} = 0$ whenever $p^j > 0$ and $y^{ji} \in Y_{ji}$ whenever $p^j = 0$. In other words, everybody will keep whatever quantities of a good with positive exchange value they have and do not need for consumption to trade for other goods. Throwing them out would only decrease their means to fund their consumption. For goods with no exchange value, in contrast, they will, as a whole, dispose of any quantities they do not need for consumption. As a result, accounting for goods disposal does not change anybody's budget set $\beta_i$. Nor does it violate *Walras's Law*.

Since every equilibrium list of exchange values is non-negative and different from zero, i.e. $p^* > 0$, it can be normalised as an element of the simplex $P$. Each exchange value on the list must be divided by the sum of all the exchange values that make up that list.

For every $p \in R_+^l \setminus \{0\}$, $\frac{1}{\sum_{j=1}^l p^j} p \in P$.

**Proof**

Firstly, $\frac{1}{\sum_{j=1}^l p^j} p > 0$ since $p > 0$ and $\sum_{j=1}^l p^j > 0$ for all $p \in R_+^l \setminus \{0\}$.

Secondly, $\sum_{j=1}^l \frac{1}{\sum_{j=1}^l p^j} p^j = \frac{1}{\sum_{j=1}^l p^j} \sum_{j=1}^l p^j = 1$

Moreover, normalising all lists of exchange values so that they become elements of the simplex $P$ does not change any consumption decisions because everybody's budget set $\beta_i$ remains unchanged. In other words, the budget set $\beta_i$ is homogenous of degree zero in the exchange values of goods. Nor does it change any disposal decisions.

**Proof**

Let $p \in R_+^l$ and $\alpha > 0$.

Firstly, $p \cdot y^i = 0$, so $\alpha p \cdot y^i = 0\alpha = 0$.

Secondly, $p \cdot x^i \leq p \cdot h^i$ for any $x^i \in \beta_i(p)$. Since $\alpha > 0$ by hypothesis, $\alpha p \cdot x^i \leq \alpha p \cdot h^i$. Moreover, for any $x$ such that $p \cdot x > p \cdot h^i$, it follows that $\alpha p \cdot x > \alpha p \cdot h^i$.

Therefore, $\beta_i(\alpha p) = \beta_i(p)$.

Finally, from the definitions of $\eta_i$ and $\xi_i$, it follows that for all $p \in R^l_+$, $\eta_i(\alpha p) = \eta_i(p)$ and $\xi_i(\alpha p) = \xi_i(p)$.

The above result means that individual decisions depend on the exchange values of goods relative to one another, not on their nominal values. Therefore, it is legitimate to restrict our attention to the simplex $P$ as long as all equilibrium lists of exchange values are indeed non-negative and the exchange values of goods are not all zero, that is, as long as $p^* > 0$.

Finally, the assumptions of local non-satiation and semi-strict convexity of preferences ensure that every fixed point yields an optimal consumption for all individuals given their budget sets at the resulting list of exchange values. Put differently, both assumptions together guarantee that every fixed point is a general equilibrium from an economic point of view. The reason is that they make the problem of maximising preferences subject to a budget constraint the equivalent of minimising the cost of achieving the associated level of utility.[44]

The image of the excess consumption correspondence at any equilibrium list of exchange values $p^*$ is strictly contained within set $Z$. That is, $\zeta(p^*) \subset Z$. Moreover, $Z$ is non-empty and chosen to be compact and convex. Consider a consumption bundle $x^i$ that an individual strictly prefers over the one allocated to them at the fixed point, i.e. $x^i \succ_i x^{i*}$. Assume that they can afford this bundle, so $x^i \in \beta_i(p^*)$. Because of the semi-strict convexity of preferences, the individual will strictly prefer any convex combination of both bundles to the one allocated to them. Formally, $x^{i*} \prec_i (1-\alpha)x^{i*} + \alpha x^i$ for all $\alpha \in\, ]0,1]$. The bundles close enough to the one allocated to them will give rise to excess consumptions contained in $Z$, i.e. $z = (1-\alpha)x^{i*} + \alpha x^i + \sum_{\substack{k=1 \\ k \neq i}}^{m} x^{k*} - \sum_{i=1}^{m} h^i \in Z$ for all $\alpha$ small enough. However, the fixed point has its excess consumption coordinates in $Z$, i.e. $z^* \in Z$. This contradicts the claim that at the equilibrium list of exchange values the individual could afford a consumption bundle that they prefer to the one they had allocated at the fixed point. In other words, $x^i \notin \beta_i(p^*)$ for all $x^i \succ_i x^{i*}$.[45]

---

[44] See Debreu (1959, p. 71) and Arrow and Hahn (1971, p. 81). On the other hand, the preference relation of individuals and their budget sets have also been assumed continuous. These two assumptions make the problem of minimising the cost of achieving a certain level of utility the equivalent of maximising preferences given a budget constraint equal to this cost (Arrow & Hahn, 1971, p. 81; Debreu, 1959, p. 69).

Moreover, the assumptions of local non-satiation and semi-strict convexity of preferences imply that everybody satisfies their budget constraint with equality at every list of exchange values.[46] Formally, for every individual $i$, $p \cdot x^i = p \cdot h^i$ for all $x^i \in \xi_i(p)$ and all $p \in P$. As a result, *Walras's Law* is satisfied with equality. In symbols, $p \cdot \zeta(p) = 0$ for all $p \in P$. Since every equilibrium list of exchange values is non-negative but not zero - $p^* > 0$ - this implies that in equilibrium, excess consumption is either zero or only negative for some goods if it is negative. That is, either $z^* = 0$ or $z^* < 0$. If $p^{j*} > 0$, then $z^{j*} = 0$ or, equivalently, $\sum_{i=1}^{m} x^{ji*} = \sum_{i=1}^{m} h^{ji}$. $z^{j*} \leq 0$, which means that $\sum_{i=1}^{m} x^{ji*} \leq$ $\sum_{i=1}^{m} h^{ji}$, only if $p^{j*} = 0$. In equilibrium, the total consumption of goods with positive exchange value certainly matches the total stocks of these goods. For those goods with no exchange value, in contrast, total consumption may well fall short of the total stocks available.

---

[45] Arrow and Hahn (1971) use the equivalence between utility maximisation subject to a budget constraint and cost minimisation to achieve a certain level of utility to prove the existence of equilibrium in terms of the latter. They prove the existence of what they call a compensated equilibrium and then add sufficient conditions to guarantee that compensated equilibria are competitive equilibria. However, if either the preferences of individuals or their budget sets are not continuous, the link between compensated and competitive equilibria breaks down.

[46] See Debreu (1959, p. 71).

## 5. Conclusion

General Equilibrium Theory proves that at least one state of general equilibrium always exists. Regardless of the initial conditions, there is always some overall consumption plan that could be fulfilled and everyone would agree on. In its most general approach, General Equilibrium Theory uses fixed-point theorems to prove this claim. As a measure of disequilibrium, it relies on the excess consumption correspondence, which is the difference between total consumption and the existing stocks of goods. It also assumes that the number of goods and economic agents is finite and that everyone takes the exchange values of goods as given and behaves rationally.

To satisfy the conditions that guarantee the existence of a fixed point, three assumptions must be made on the excess consumption correspondence and one assumption on all equilibrium lists of exchange values. Firstly, the excess consumption correspondence is upper hemicontinuous, with each point in the range of this correspondence having as many coordinates as the total number of goods. Secondly, the excess consumption correspondence is convex valued. If the difference between total consumption at a given list of exchange values and the total stocks of goods can be two different points, then it is also every point in between. Thirdly, to guarantee that the fixed point is an equilibrium from an economic point of view, the correspondence must satisfy *Walras's Law*. Therefore, the exchange value of total consumption never exceeds the exchange value of the total stocks of goods. Lastly, no equilibrium list of exchange values can be zero for all goods, which means that goods cannot all be free in equilibrium.

Seven assumptions must be made to ensure that the excess consumption correspondence is upper hemicontinuous from individual behaviour. The set of possible consumptions for each individual is closed, convex and bounded from below. Individuals maximise their preferences and preferences are closed. The total stocks of goods are available to anyone who can afford them and hence are given quantities in the system of exchange. Lastly, the exchange value of everyone's initial goods endowment is larger than that of their cheapest possible consumption at every list of exchange values.

The assumptions that an individual's set of possible consumptions is closed and bounded from below mean that their set of feasible consumptions given the available resources, assumed finite, is compact. The assumption that the individual's set of possible consumptions is convex makes every point in their consumption set connected to every other by a straight line between the points. It also implies that goods are perfectly divisible. These three assumptions, coupled with the assumption that the exchange value of their initial goods endowment is larger than that of their cheapest possible consumption at every list of exchange values, make their budget set continuous on their feasible consumption set. If the exchange values of goods change continuously, so does the set of alternative consumptions that an individual can afford.

The assumption that individual preferences are closed and the assumption that their consumption set is convex and hence connected allows representing each individual's preferences by a continuous utility function. The fact that individuals maximise a continuous utility function, coupled with their budget set being continuous on their feasible consumption set, makes an individual's consumption upper hemicontinuous on this set. As a result, total consumption is upper hemicontinuous since it is the sum of the consumption correspondences of individuals.

The assumption that the available stocks of goods are constant, coupled with total consumption being upper hemicontinuous, makes the excess consumption upper hemicontinuous. The fact that everybody faces a budget constraint, in turn, ensures that the excess consumption correspondence satisfies *Walras's Law*.

To guarantee that the excess consumption correspondence is convex valued, it is assumed that individual preferences are convex. If there are two consumption bundles such that the individual prefers one to the other or is indifferent between them, they will also prefer any bundle in between to the latter or be indifferent between them. As the budget set of an individual is also convex, this makes everybody's consumption correspondence convex at every list of exchange values. The reason is that everyone's consumption correspondence is the outcome of their maximising their preferences within their budget sets. Consumers might have a collection of alternative optimal consumption plans at any such list between which they are indifferent rather than just one.

To ensure that no equilibrium list of exchange values is zero for all goods, it is assumed that individuals are insatiable over all feasible allocations of the total stocks of goods (local non-satiation). The assumption fulfils its purpose by creating a context of scarcity. It also guarantees that the excess consumption correspondence yields an optimal consumption at every fixed point for each individual. The reason is that individual preferences are not just convex but semi-strictly convex. If there are two consumption bundles, one of which an individual strictly prefers to the other, they will strictly prefer every bundle in between to the latter. Therefore, at an equilibrium list of exchange values nobody can afford any consumption bundle they prefer to those they receive at the state of general equilibrium associated with that list. To guarantee that no exchange value is negative in equilibrium, it is further assumed that any quantities of goods that nobody needs can be disposed of for free (free disposal).

The assumptions of local non-satiation and semi-strict convexity of preferences also imply that everyone only just meets their budget constraint at every list of exchange values. As a result, the exchange value of total consumption always matches that of the total stocks of goods. Since the exchange values of goods are non-negative but not all zero in equilibrium, excess consumption is either zero or only negative for some goods in equilibrium. Total consumption matches the total stocks of those goods with positive exchange value but may fall short of these stocks for free goods.